\documentclass[times,twocolumn]{aastex62}

\usepackage{CJK}
\usepackage{amsmath}
\usepackage{multirow}
\usepackage{cases}
\usepackage{graphicx}
\usepackage{subfigure}
\usepackage{natbib}
\usepackage{color}
\usepackage{tabularx}
\usepackage{bm}
\usepackage{threeparttable}
\usepackage{gensymb}

\newcommand{\HL}[1]{\textcolor{black}{#1}}
\newcommand{\thetae}{\theta_{\rm E}}
\newcommand{\thetas}{\theta_{*}}

\newcommand{\te}{t_{\rm E}}
\newcommand{\Ds}{D_{\rm S}}
\newcommand{\Dsi}{D_{{\rm S},i}}
\newcommand{\Dl}{D_{\rm L}}

\newcommand{\Rs}{R_{\rm S}}
\newcommand{\Rsi}{R_{{\rm S},i}}

\newcommand{\Mearth}{M_{\oplus}}

\DeclareGraphicsExtensions{.pdf,.png,.jpg}

\shorttitle{}
\shortauthors{Yang et al.}

\begin{document}
\begin{CJK*}{UTF8}{gbsn}
\title{{\large How Rare are TESS Free-Floating Planets?}}

\correspondingauthor{Hongjing Yang, Weicheng Zang}
\email{hongjing.yang@qq.com, 3130102785@zju.edu.cn}

\author[0000-0003-0626-8465]{Hongjing Yang (杨弘靖)}
\affiliation{Department of Astronomy, Tsinghua University, Beijing 100084, China}

\author[0000-0001-6000-3463]{Weicheng Zang (臧伟呈)}
\affiliation{Center for Astrophysics $|$ Harvard \& Smithsonian, 60 Garden St.,Cambridge, MA 02138, USA}
\affiliation{Department of Astronomy, Tsinghua University, Beijing 100084, China}

\author[0000-0002-4503-9705]{Tianjun Gan (干天君)}
\affiliation{Department of Astronomy, Tsinghua University, Beijing 100084, China}

\author[0000-0003-2337-0533]{Renkun Kuang (匡仁昆)}
\affiliation{Department of Astronomy, Tsinghua University, Beijing 100084, China}

\author{Andrew Gould} 
\affiliation{Max-Planck-Institute for Astronomy, K\"onigstuhl 17, 69117 Heidelberg, Germany}
\affiliation{Department of Astronomy, Ohio State University, 140 W. 18th Ave., Columbus, OH 43210, USA}

\author[0000-0001-8317-2788]{Shude Mao (毛淑德)}
\affiliation{Department of Astronomy, Tsinghua University, Beijing 100084, China}

\begin{abstract}
Recently, Kunimoto et al. claimed that a short-lived signal in the Transiting Exoplanet Survey Satellite (TESS) Sector 61 database was \HL{possibly} caused by a microlensing event with a terrestrial-mass free-floating planet (FFP) lens. In this study, we investigate TESS's ability to detect microlensing FFPs by considering the detailed source information (e.g., distance and radius), the TESS photometric accuracy, and finite-source effects. Using the FFP mass function from microlensing surveys toward the Galactic bulge, we find that only $0.0018$ microlensing events are expected to be detected in TESS Sector 61 for the entire planetary mass range. The reported signal is unlikely to be a real microlensing event, which is consistent with the evidence from the long-term OGLE data that the signal was likely due to a stellar flare. By extrapolating our result to fainter stars until $T = 16$ mag and adopting a possible optimized search algorithm, we find that only $\sim 1$ FFP events can be detected in the entire TESS mission within the first 7 years. Significant improvments of our understanding of FFPs still requires future satellite missions, such as Roman and Earth 2.0, which can detect thousands of FFPs.
\end{abstract}

\section{Introduction}\label{sec:intro}

Although deep high-resolution imaging is capable of finding Jupiter-mass free-floating planets (FFPs), e.g., Jupiter-Mass Binary Objects (JuMBOs, \citealt{JuMBOs}), the gravitational microlensing technique \citep{Einstein1936,Paczynski1986} is the only method that can explore FFPs in the entire planetary-mass range. Although limited by systematics in the data and small number of statistics, \cite{Sumi2011} opened the field of microlensing FFPs by studying 474 microlensing events observed by the Microlensing Observations in Astrophysics (MOA, \citealt{Sako2008}) group and claimed about two Jupiter-mass FFPs per star. The large population of Jupiter-mass FFPs was later excluded by the larger samples from the Optical Gravitational Lensing Experiment (OGLE, \citealt{OGLEIV,Mroz2017a}), the Korean Microlensing Telescope Network (KMTNet, \citealt{KMT2016,Gould2022_FFP_EinsteinDesert}), and the MOA group itself \citep{Naoki_FFP,Sumi2023_FFPMF}. Nevertheless, the larger samples found dozens of events with the extremely short Einstein timescale ($\te < 0.5$ days), of which \HL{nine} \citep{MrozNeptune,Mroz2FFP,OB190551,OB161928,KB192073,KB172820,Naoki_FFP,KB232669_Jung2024} have an angular Einstein radius, 
\begin{equation}\label{equ:thetae}
    \thetae = 4.946 \sqrt{\frac{\Ds}{\Dl}-1} \left(\frac{M_{\rm L}}{\Mearth}\right)^{\frac{1}{2}} \left(\frac{\Ds}{\rm kpc}\right)^{-\frac{1}{2}} {\rm \mu as},
\end{equation}
below the Einstein desert ($9~\mu{\rm as} < \thetae < 26~\mu{\rm as}$, \citealt{Gould2022_FFP_EinsteinDesert}). Here $\Ds$ and $\Dl$ are the source and lens distances, and $M_{\rm L}$ is the lens mass. These events were thus probably caused by FFPs with masses from Mars mass to Neptune mass, leading to two studies of the mass function of FFPs \citep{Gould2022_FFP_EinsteinDesert, Sumi2023_FFPMF}. 

The duration of the microlensing FFP events is short (i.e., $\lesssim 1$ days), even considering finite-source (FS) effects \citep{1994ApJ...421L..75G,Shude1994,Nemiroff1994} caused by giant sources and the event rate of microlensing FFPs is low, with $\Gamma \lesssim 10^{-8}~{\rm yr}^{-1}~{\rm star}^{-1}$. Furthermore, the microlensing effect is unpredictable and unrepeatable. Therefore, high-cadence (e.g., $\geq 1~{\rm hr}^{-1}$) large-area surveys are needed. The OGLE, MOA, KMTNet and the PRime-focus Infrared Microlensing Experiment (PRIME, \citealt{PRIME_pre}) are conducting high-cadence large-area microlensing surveys toward the Galactic bulge, but due to weather, Moon, and the diurnal and annual
cycles, the detection rate for FFPs from these surveys is still low, with $\sim 1$ event per year. 

High-cadence large-area surveys from space-based telescopes can overcome the difficulties on the ground. Detecting FFPs is one of the primary scientific objectives of both the Nancy Grace Roman Space Telescope \citep[{\it Roman}, former {\it WFIRST}, ][]{Spergel2015, MatthewWFIRSTI, Johnson2020} and the Earth 2.0 mission \citep{ET}, and they will detect ${\cal O}(10^3)$ FFPs \citep{Sumi2023_FFPMF}. In addition, the Chinese Space Station Telescope ({\it CSST}, \citealt{CSST_Wei}) and the {\it Euclid} satellite \citep{Bachelet2022} can improve the detection and mass measurements for FFPs from the satellite microlensing parallax \citep{1966MNRAS.134..315R, 1994ApJ...421L..75G, Gould1995single}. 

Recently, \cite{TESS_FFP} claimed the detection of a \HL{candidate} terrestrial-mass microlensing FFP event on the TIC-107150013 source star using the Sector 61 (hereafter S61) data of the Transiting Exoplanet Survey Satellite (TESS, \citealt{Ricker2015}), Later a subsequent work using the OGLE data suggested that the short-lived signal was likely due to a stellar flare \citep{MrozTESSFFP}. Based on this detection and a rough estimate of the TESS yields, \cite{TESS_FFP} also claimed that TESS has the opportunity to significantly improve our understanding of FFPs with terrestrial and sub-terrestrial masses. In this paper, we estimate TESS's ability to detect microlensing FFPs and thus assess the authenticity of that detection by considering the detailed source information (e.g., distance and radius), the TESS photometric accuracy, and FS effects, which were neglected by the estimate of \cite{TESS_FFP}. 

The paper is structured as follows. In Section \ref{sec:method}, we introduce our methodology to estimate TESS's detection efficiency and the Galactic model we used. Then, the resulting TESS expected yields of FFPs are presented in Section \ref{sec:res}. Finally, we discuss the possible errors of our estimate and possible optimizations for TESS FFP searches in Section \ref{sec:dis}.

\section{Methodology}\label{sec:method}
The number of observed microlensing events depends on two factors, the intrinsic rate of microlensing events and the fraction of these events that is detectable under specific observation conditions, such as the cadence, baseline length, and noise level. These two effects are entangled in the calculation, which will be explained accordingly. We consider both the properties of the source stars and the actual noise level present in TESS data.

For an individual source star (hereafter denoted by subscript $i$) with a distance of $\Dsi$ and a specific lens mass $M_{\rm L}$, the microlensing event rate can be written as \citep[e.g.,][]{CMST}
\begin{equation}
\begin{split}
    \Gamma_i = \int &dM_{\rm L}\times \\ 
    &\int_0^{\Dsi} d D_{\rm L} \sigma_i(M_{\rm L}, D_{\rm L}) D_{\rm L}^{2} n(M_{\rm L},D_{\rm L}) \langle\mu_{\rm rel}(D_{\rm L})\rangle,
    \label{eq:eventrate}
\end{split}
\end{equation}
where $\sigma_i(M_{\rm L}, D_{\rm L})$ denotes the angular cross-section, $n(M_{\rm L}, D_{\rm L})$ is the number density of lens objects at a specific mass and distance, and $\langle\mu_{\rm rel}(\Dl)\rangle$ represents the mean lens-source relative proper motion at a given distance.  The unit of $\Gamma_i$ is events per year. Therefore, the expected number of detections for a given observation baseline $T_{\rm obs}$ is 
\begin{equation}
    N_{\rm FFP} = T_{\rm obs} \sum_i^{N_{\star}} \Gamma_i,
\end{equation}
where $N_{\star}$ is the total number of stars monitored.
Three crucial elements in Eq. (\ref{eq:eventrate}) determine the event rate, namely $\sigma_i(M_{\rm L}, D_{\rm L})$, $n(M_{\rm L},D_{\rm L})$, and $\langle\mu_{\rm rel}(D_{\rm L})\rangle$. We individually discuss each element below.

The first element is the cross-section $\sigma_i(M_{\rm L}, D_{\rm L})$, which determines the area on the sky where a lens can cause a detectable microlensing event, and we define it as 
\begin{equation}
   \sigma_i(M_{\rm L}, D_{\rm L}) = 2 u_{T,i} \thetae,
\end{equation}
where $u_{T,i}$ is the maximum impact parameter for an event that satisfies the detection threshold. For a given $(M_{\rm L}, D_{\rm L}, D_{\rm S})$, $\thetae$ can be derived by Eq. \ref{equ:thetae}. To derive $u_{T,i}$ of each source $i$, we obtain the TESS magnitude $T_i$, distance $\Dsi$, and stellar radius $\Rsi$ from the TESS Input Catalog v8 \citep[TIC,][]{Stassun2019_TIC}, yielding the source radius normalized to the angular Einstein radius as
\begin{equation}
    \rho=\theta_*/\thetae,
\end{equation}
where $\theta_* = \Rsi/\Dsi$ is the angular source radius, and thus the maximum magnification $A_{i, \rm L, max}(\rho)$ as a function of $\rho$ for a given lens. We calculate $A_{i, \rm L, max}(\rho)$ using the \texttt{pyLIMA} lensing model package \citep{pylima}. Then, we estimate the noise level $\sigma_{N,i}$ in units of parts-per-million (ppm) using software \texttt{ticgen}\footnote{\url{https://github.com/tessgi/ticgen}} \citep{ticgen, Stassun2018_TIC_ticgen}, where the integration time 200~s for Sector 61 is adopted. In addition, we determine the minimum magnification $A_{T,i}$ required to achieve a signal-to-noise ratio (SNR) exceeding 10 \citep[defined by the FFP search algorithm of][]{TESS_FFP} using
\begin{equation}\label{eq:A_thres}
    A_{T,i} = 1 + 10 \sigma_{N,i}.
\end{equation}
If $A_{T,i} \geq A_{i, \rm L, max}(\rho)$, the microlensing events from the given lens-source configuration are undetectable and we have $u_{T,i} = 0$. If $A_{T,i} < A_{i, \rm L, max}(\rho)$, we derive $u_{T,i}$ by $A_{i, \rm L}(\rho, u_{T,i}) = A_{T,i} $ using the \texttt{pyLIMA} package. 
\HL{Note that we require that each point on the peak has a signal-to-noise ratio (SNR) larger than 10, which is different from the threshold adopted by \citet{TESS_FFP} that uses the SNR over the entire event.
Because microlensing events occur only once, they are relatively sensitive to short-timescale systematic errors. 
Our stricter threshold can help overcome these systematic errors and help better characterize the physical parameters.
A detailed discussion about the criteria will be presented in Section \ref{sec:snr}.
}

The second element is the lens number density $n(M_{\rm L}, D_{\rm L})$. We assume that the density profile is independent of lens mass $M_{\rm L}$, i.e., $n(M_{\rm L}, D_{\rm L})=f(M_{\rm L})n(D_{\rm L})$. Here $f(M_{\rm L})=d N_{\rm lens}(M_{\rm L})/d M_{\rm L}$ represents a scaling factor of the relative abundance of lenses with a specific mass $M_{\rm L}$ compared to the total number of main-sequence stars.
For the mass-independent density profile $n(D_{\rm L})$, we follow the model described in \citet{Yang2021_GalacticModel}.
Given that TESS sources are mostly nearby stars and located in the Galactic disk, we adopt the exponential disk profile
\begin{equation}
    n(R,z) = n_0 e^{\left(-\frac{R-R_0}{R_d}-\frac{|z|}{z_d}\right)},
\end{equation}
where $R$ and $z$ are radial and vertical distances to the Galactic center in Galactic cylindrical coordinates. The scale length and scale height of the disk $(R_d, z_d)$ are set to $(2.5, 0.325)~{\rm kpc}$ \citep{BinneyTremaine2008_GalacticDynamics}. 
The normalization factor $n_0$ is derived from the local stellar density in the solar neighborhood, $0.14$~pc$^{-3}$. For each source star, $n(D_{\rm L})$ is evaluated along its line-of-sight. 

The last crucial element is the mean relative proper motion, $\langle\mu(D_{\rm L})\rangle$, which reflects the projected velocity difference between the lens and source star onto the viewing plane. While the actual velocity distribution is complex, involving Galactic rotation and local velocity dispersion, we simplify it as follows. We take a constant velocity $V=55~{\rm km/s}$ with 3-dimensional random directions for both the sources and lenses. More specifically, for a given source and lens distances $(\Ds, D_{\rm L})$, we sample 100 velocity pairs $(\vec{V}_{{\rm S},j}, \vec{V}_{{\rm L},j})$ 
to estimate the mean relative proper motion
\begin{equation}
    \langle\mu_{\rm rel}(D_{\rm L})\rangle = \frac{1}{N_{\rm sample}} \sum_j^{N_{\rm sample}}\left| \frac{\vec{V}_{{\rm L},j,\perp}}{D_{\rm L}} - \frac{\vec{V}_{{\rm S},j,\perp}}{\Ds} \right|,
\end{equation}
where $\vec{V}_{{\rm S},j,\perp}$ and $\vec{V}_{{\rm L},j,\perp}$ denote the source and lens velocities projected onto the viewing plane, respectively. 

In addition, cadence is usually considered in microlensing sensitivity calculations. However, the high cadence of TESS S61 (200~s) ensures sufficient sampling (i.e., $>5$ data points achieve a SNR $>10$) because the source self-crossing time $t_* =  \rho\te > 1$ hr for all $T < 13.5$ stars even with an extreme $\mu_{\rm rel} = 50 ~\mathrm{mas\ yr^{-1}}$.

\section{Expected yields of TESS}\label{sec:res}

\subsection{The Sector 61}

\begin{figure}
    \centering
    \includegraphics[width=\columnwidth]{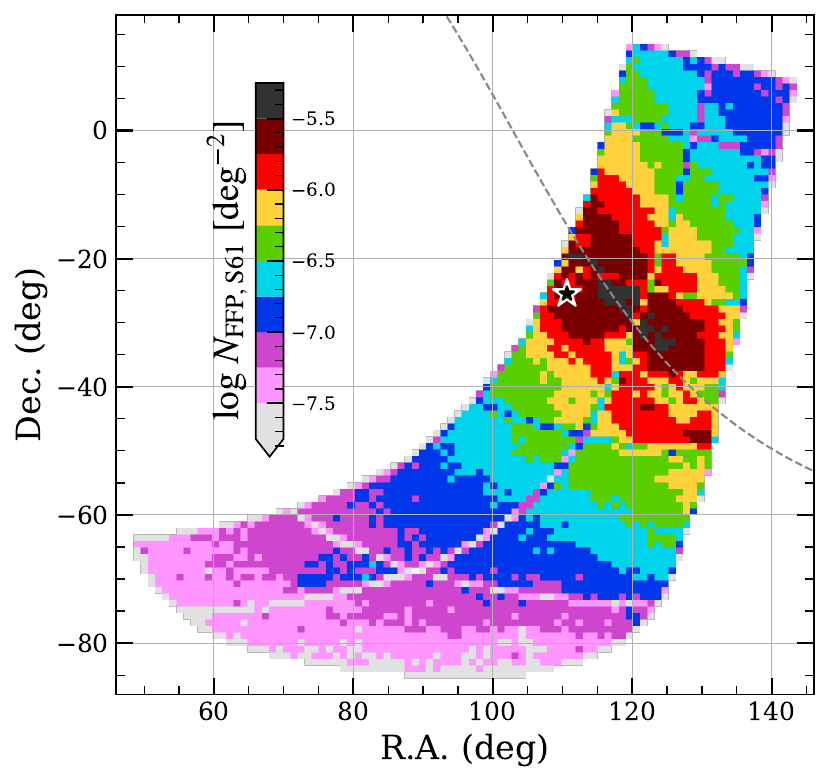}
    \caption{The expected free-floating planet yields of TESS Sector 61 in sky region bins. The bin sizes are $\Delta {\rm R.A.} = \Delta{\rm Dec.} = 1^\circ$.
    The integral over the entire Sector 61 gives $N_{\rm FFP, S61}=1.81\times10^{-3}$. 
    The dashed line represents the Galactic plane. The ``$\star$'' symbol marks TIC-107150013, the candidate event reported by \citet{TESS_FFP}.
    }
    \label{fig:er_sky}
\end{figure}

To maintain consistency with the systematic search conducted in \cite{TESS_FFP}, we retrieved all stars within the TESS S61 field that have light curves produced by the Quick-Look Pipeline \citep[QLP,][]{Huang2020_QLP,Kunimoto2021} from TIC. This resulted in a total of 1,288,149 stars included in our analysis.

We adopt the mass function of FFPs measured by \cite{Sumi2023_FFPMF}, which has
\begin{equation}
    f(M)=\frac{dN_{\rm lens}(M)}{d\log M} = 2.18 \left(\frac{M}{8 \Mearth}\right)^{-0.96}~{\rm star}^{-1},
\end{equation}
and is consistent with the mass function of \cite{Gould2022_FFP_EinsteinDesert}. We divide the FFP mass range into seven uniform logarithmic bins, each spanning 1 dex in mass, namely $\log (M/\Mearth) =(-3,-2,-1,0,1,2,3)$. These bins correspond to FFP populations of 
($1.48\times10^4$, $1.62\times10^3$, $1.78\times10^2$, $1.95\times10^1$, $2.14\times10^0$, $2.35\times10^{-1}$, $2.57\times10^{-2}$) per star. 

To obtain $\Gamma_i$ of each star, we numerically integrate Eq. \ref{eq:eventrate} by considering different masses of FPPs and uniformly sampling 10 values\footnote{We have tried more samples but the resulting expected TESS yields are the same because the number of the input source stars is large enough to eliminate the fluctuations in sampling.} of the lens distance $D_{\rm L}$ from 0 to $\Dsi$. Figure \ref{fig:er_sky} shows the expected number of FFP detections across the S61 field, considering the effective observation baseline, $T_{\rm obs}=23.6~{\rm d}$, of S61 \citep{TESS_FFP}. The expected number of FFP events is highest on the Galactic plane and decreases along the Galactic latitude because of the distribution of the stellar density (i.e., the highest on the Galactic plane).

By summing the individual expected number of FFP detections across all stars, we obtain the total expected FFP detections of S61 as
\begin{equation}
    N_{\rm FFP, S61} = 1.81 \times 10^{-3},
\end{equation}
which indicates that the possibility of detection of FFP in the S61 data is low. In addition, the long-term OGLE data show that this signal was likely due to a stellar flare \citep{MrozTESSFFP}. Moreover, he found similar symmetric flares in other stars. Combining the two facts, we conclude that the short-lived signal reported by \cite{TESS_FFP} is unlikely to be a real microlensing event. 

\begin{figure}
    \centering
    \includegraphics[width=\columnwidth]{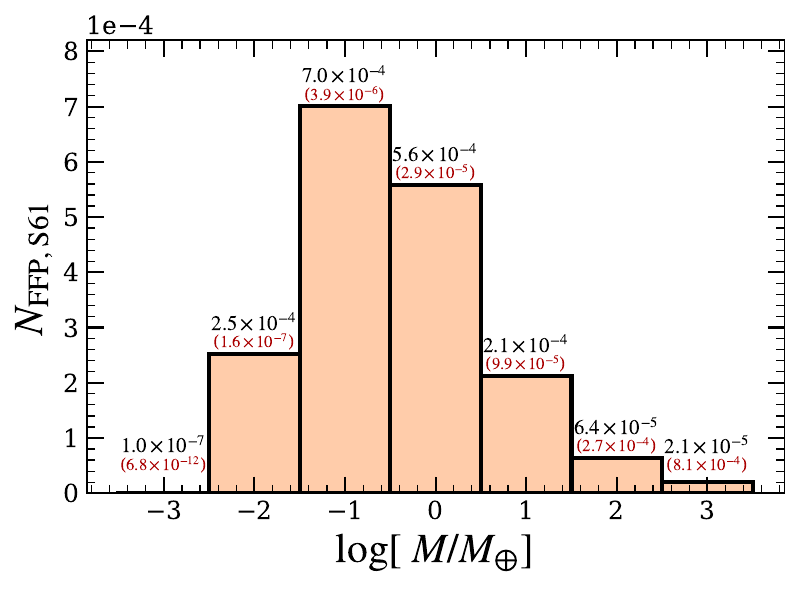}
    \caption{The expected free-floating planet yields of TESS Sector 61 in mass bins. Each mass bin represents 1 dex in mass. \HL{On the top of each bin, the sensitivity (red) and the expected number (black) after considering the FFP mass function from \citet{Sumi2023_FFPMF} are labeled.} Summing over all mass bins gives \HL{the total expected number} $N_{\rm FFP, S61}=1.81\times10^{-3}$.}
    \label{fig:er_mass}
\end{figure}

In addition, our simulation of the expected TESS yield in FFPs has two main differences from the estimate made by \cite{TESS_FFP}. First, as shown in Figure \ref{fig:er_mass}, our simulation suggests that TESS is most sensitive to FFPs with masses of $0.1\Mearth \lesssim M \lesssim 1\Mearth$, while \cite{TESS_FFP} claim that the TESS FFPs are expected to be largely due to FFPs with masses of $0.01\Mearth \lesssim M \lesssim 0.1\Mearth$. Second, compared to the estimate using Eq. (5) of \cite{TESS_FFP}, our simulation shows $\sim 30$ times lower expected yields for $M \sim 0.1\Mearth$ and $\sim 10$ times lower for $M \sim 1\Mearth$. 

\begin{figure}
    \centering
    \includegraphics[width=\columnwidth]{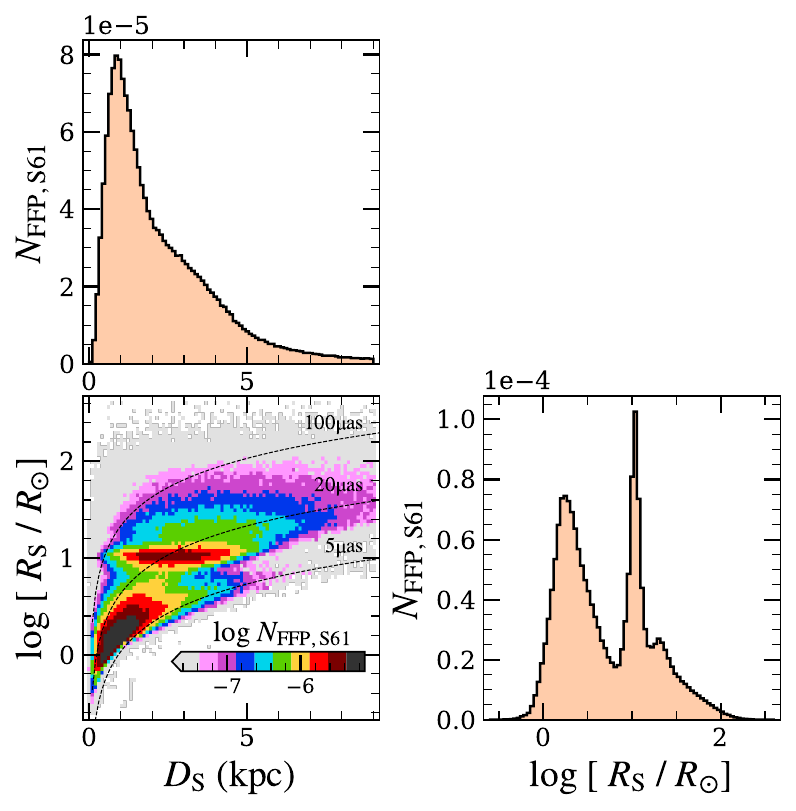}
    \caption{The expected free-floating planet yields of TESS Sector 61 in source distance and stellar radius bins. The bin sizes are $\Delta\Ds=0.1~{\rm kpc}$ and $\Delta\log[\Rs/R_{\odot}]=0.1$.
    The black dashed lines are the contours of the angular stellar radius $\thetas$.}
    \label{fig:er_DsRs}
\end{figure} 

Most of the difference was due to that \cite{TESS_FFP} did not consider the actual TESS photometric accuracy and FS effects. Figure \ref{fig:er_DsRs} displays the expected TESS FFP yields across different source distances and stellar radii, from which most of the sensitivity is from main-sequence stars within 2 kpc and red giants with a stellar radius of $\sim 10R_{\odot}$ in a wider distance range. Almost all of the sources have $\theta_* > 5~{\rm \mu as}$ and while $\thetae \lesssim 5~{\rm \mu as}$ for $M < 1\Mearth$ according to Eq. (\ref{equ:thetae}), thus the light-curves of $M < 1\Mearth$ should be dominated by FS effects and the maximum magnification is suppressed and related to the FFP mass by \citep{Shude1994,Gould_FS}
\begin{equation}
    A_{\rm max} (\rho) - 1  \sim \frac{\HL{2}}{\rho^2} = \frac{2\thetae^2}{\theta_*^2} \propto M. 
\end{equation}
Combined with the actual TESS photometric accuracy, most $u_T$ are smaller than 5, which was adopted by \cite{TESS_FFP} for the cross-section, and even $u_T = 0$ for low-mass FFPs with a close lens-source distance. 

This can be quickly confirmed by a visual inspection from the light curve of TIC-107150013 shown in \cite{TESS_FFP}. The photometric scattering of TESS data is $\sim 0.006$ mag ($\sim 600$~ppm). The event has $\rho\sim4.5$ and $\thetae\sim4~{\rm \mu as}$. 
Thus, for most of $M < 1\Mearth$ FFPs with $\thetae < 2~{\rm \mu as}$ (therefore $\rho>9$), the TIC-107150013 star can only have a maximum magnification change of $A_{\rm max}-1<0.025$, which would not meet the criterion of \cite{TESS_FFP} that SNR $>10$ for $>5$ data points.

\subsection{The Entire TESS Mission}\label{sec:entire}

Here we extend our estimate to the entire TESS mission. The TESS mission can be divided into three phases. The first is the primary mission, including 2 cycles (26 sectors) during which the full-frame images were stacked every 30~min.  After the primary mission, TESS operated its first extended mission for 2 more cycles (29 sectors). In this phase, the full-frame images were taken every 10~min. After that, the secondary extended mission has been started and is currently ongoing. In this phase, three additional cycles including 41 sectors will obtain the full-frame images with a cadence of 200~s. The major difference between these phases is the cadence or the integration time. Here we still use the catalog from S61, but incorporate the SNR from the different integration times, 30~min and 10~min, for the primary and first extended mission, respectively. We also account for the increased probability of missing signals due to lower cadences compared to S61.

We find the primary mission and the first extended mission are expected to yield $3.68\times10^{-3}$ and $2.85\times10^{-3}$ FFPs per sector, respectively. These values are higher than S61, primarily due to the higher SNR from the longer integration time. In the mean time, the cadences are still sufficient for the FFP detection. We note that S61 has 35\% more TIC stars than the average of Sectors 1 -- 70 \citep{TESS_FFP}. Considering a correction factor of 1/(1+35\%) = 74\%, the total expected FFP yield is $N_{\rm FFP, 7yr}=0.2$ for all 96 sectors of the first seven cycles of the TESS mission.

\section{Discussion}\label{sec:dis}
\subsection{\HL{Definition of the Signal-to-Noise Ratio}}
\label{sec:snr}
\HL{To improve our knowledge of the FFP populations, the ultimate goal of the FFP-like light curve search is to well characterize the events, in other words, to well constrain the parameters including the Einstein timescale $t_{\rm E}$ and the finite-source size $\rho$. To achieve this, we adopt a different criterion with \citet{TESS_FFP}. }
\HL{\citet{TESS_FFP} use the SNR defined by the signal over the noise of the entire event (SNR$_{\rm entire}$) as the threshold of the search.
However, the SNR$_{\rm entire}>10$ threshold would not meet the above requirement. 
In comparison, our proposed criteria can result in well constrained parameters and the candidate event TIC-107150013 can still be recovered.}

\begin{figure}
    \centering
    \includegraphics[width=\columnwidth]{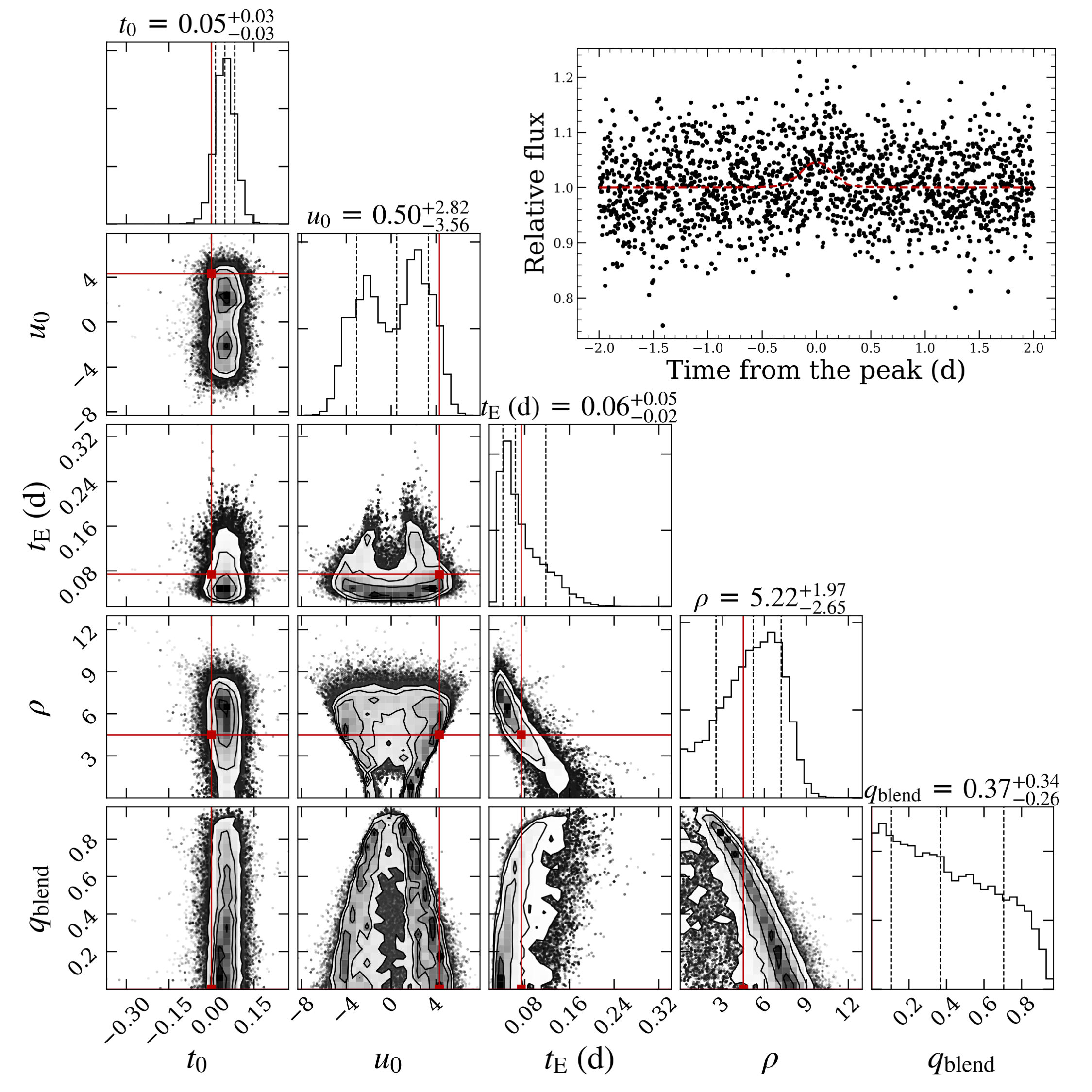}
    \caption{\HL{Light curve of a simulated SNR$_{\rm entire}>10$ but SNR$_{\rm each}\sim1$ FFP event (upper right panels) and the MCMC optimization results of it (lower left coner plot). The true (input) values are marked in red. Parameters including $u_0$, $\te$, $\rho$, and $q_{\rm blend}$ are degenerated and are not well constrained.}}
    \label{fig:lowSNR}
\end{figure}

\HL{To illustrate this, we simulate an FFP light curve similar to TIC-107150013 but enlarge the noise by a factor of 10 (from $\sim5,000$ ppm to $\sim 50,000$ ppm). In this case, the signal-to-noise ratio of each peak point is SNR$_{\rm each}\sim1$, while the SNR$_{\rm entire}>10$ is still satisfied. The upper right panel of Figure \ref{fig:lowSNR} shows the light curve and the model. We run the Markov-Chain Monte-Carlo (MCMC) optimization to fit the light curve and obtain the posterior probability distribution of the parameters. To fit the light curve, apart from $t_0$, $u_0$, $\te$ and $\rho$, an additional parameter $q_{\rm blend}$ is included that describes the blending fraction.} 

\HL{The resulting posterior is shown in the left lower corner plot of Figure \ref{fig:lowSNR}. The parameters $u_0$, $\te$, $\rho$ and $q_{\rm blend}$ are all degenerated and not well constrained. The finite-source size $\rho$ is only $\sim2\sigma$ measured and not significantly non-zero. 
For comparison, in the case of TIC-107150013 \citep{TESS_FFP}, even after adding $q_{\rm blend}$ as a free parameter, $\rho$ is still well constrained\footnote{And $q_{\rm blend}$ is well consistent with zero.}.}

\HL{In addition, our single-point SNR threshold is consistent with that in \citet{TESS_FFP} if the noise is dominated by the red noise. We will discuss the results when changing our threshold in Section \ref{sec:dis}.
Indeed, the stricter threshold could lead to a higher false positive rate. In Section \ref{sec:optimize}, we will try to optimize the threshold, but the results will not have order-of-magnitude differences.}

\subsection{Possible Errors of the Yields}\label{sec:errors}
Our calculations involve certain assumptions that could potentially induce bias or uncertainties; below we explain these separately.

The TESS Input Catalog (TIC) lacks distance and/or radius information for 92,437 stars in S61. We excluded these stars in our calculations and assumed zero sensitivity to FFP events. If we instead assign them the average sensitivity, the total expected yield would only increase by 7\% and the overall conclusions of this paper would not change.

In Section \ref{sec:method}, we adopted a simplified model for the velocity distributions of both lens and source stars. They were assumed to have constant velocities with random directions while the Galactic rotation and non-isotropic velocity dispersion were ignored.
However, we note that the velocity value we adopt, $V=55~{\rm km/s}$, is higher than the local velocity dispersion $\sim50~{\rm km/s}$ \citep{GaiaDR2_kinematics}. This assumption leads to a higher value of $\langle\mu\rangle$. This higher $\langle\mu\rangle$ could effectively mimic the influence of the Galactic rotation and non-isotropic velocity dispersion. Therefore, the final results are not sensitive to this assumption.

Blending (also called flux contamination in TIC), where multiple stars appear as a single source in the observations, was not considered when estimating the magnification threshold (Eq. \ref{eq:A_thres}). 
Given TESS's large pixel size $\sim21^{\prime\prime}$, blending can be significant \HL{(an extreme case is ``pixel lensing'' where the total flux is dominated by the blending flux, e.g., \citealt{Gould1995_pixellensing})}.  However, the current search targeting only the brighter sources (TESS magnitude $<13.5$) might not be considerably affected.
The flux contamination parameter within a subset of stars in TIC is provided by \citet{Stassun2018_TIC_ticgen,Stassun2019_TIC}. We test our hypothesis using these stars ($\sim 1/3$ out of the 1.3 million) in TIC and find a $\sim20\%$ decrease in the expected yield. 
Therefore, the qualitative results that TESS S61 is inadequate for detecting FFPs have not significantly changed. Future searches (if any) for fainter sources would require a more comprehensive sensitivity analysis that incorporates blending.

\HL{The mass function we adopted \citep{Sumi2023_FFPMF} was derived from a statistical investigation based on 9-yr MOA observations toward the Galactic bulge. The mass function was measured by correcting the actual detection numbers with the survey sensitivity. The mass function is uncertain due to the small-number detection. However, the sensitivity, i.e., how sensitive is the survey to detect certain mass FFPs is precisely calculated and independent of mass function. 
This is also true in our calculations. In Figure \ref{fig:er_mass}, we mark the sensitivity on the top of each bin (red numbers), i.e., how many FFPs can be detected for TESS S61 observations if the FFP has the same number density as main-sequence stars. These numbers are free from the mass function assumption. Furthermore, MOA has discovered $\mathcal{O}(10)$ very short events that are FFP candidates, including a strong sub-Earth FFP candidate \citep{Naoki_FFP}, while the KMTNet and OGLE surveys should have more FFP events given the survey areas, cadences, and photometric accuracy. Compared to the ground-based surveys, TESS is expected to detect only about one FFP event using the same mass function. Thus, TESS is less sensitive to FFPs compared to current ground-based surveys.}

\subsection{Optimizing Future TESS FFP Searches}\label{sec:optimize}

In Section \ref{sec:entire}, we find that a primary-mission-like observational mode of TESS is more sensitive to FFPs compared with other phases. 
\HL{This was due to the threshold we used in Section \ref{sec:method} requires $\geq5$ data points exceeding the SNR threshold of 10 and thus a longer integration time has a higher sensitivity when the cadence is sufficient. }
Thus, for S61, a potentially more effective strategy is using lower per-point SNR thresholds but increasing the required number of points, for example, SNR$>$5 and $N_{\rm pt}>20$. The revised strategy is the same as the original one in terms of $\chi^2$. We calculate the expected FFP yields for S61 using this proposed strategy and obtain an increase in the total number of detections from $1.81\times10^{-3}$ to $3.16\times10^{-3}$. Note that this improvement might be slightly overestimated as the correlated (red) noise is ignored. \HL{Moreover, this result can hardly be extrapolated to SNR$\lesssim3$, because such events would not have good parameter constraints and thus not be considered as robust events.} Nevertheless, the idea is that the search strategy could be further optimized within each observational phase.

\begin{figure}
    \centering
    \includegraphics[width=\columnwidth]{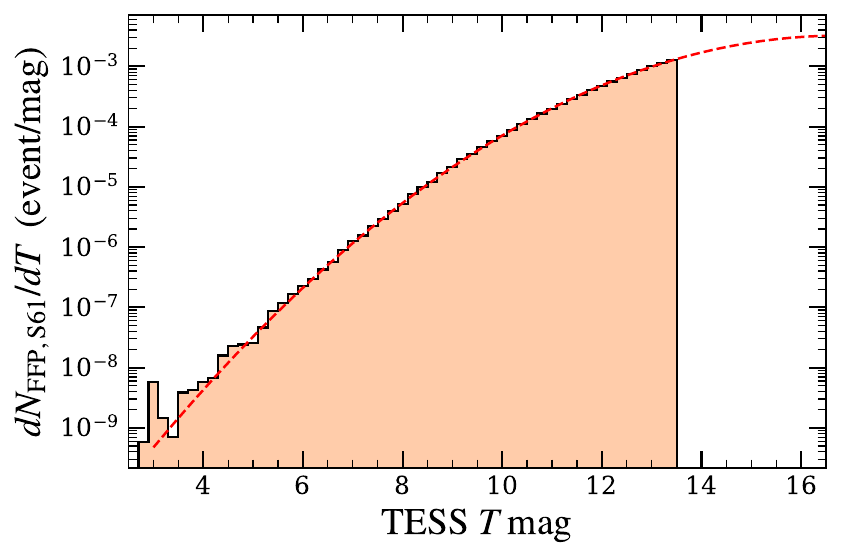}
    \caption{The expected free-floating planet yields of TESS Sector 61 as a function of the source magnitude using the proposed new strategy. The shaded region plots the expected yields and the red dashed line is a quadratic fit to its logarithm (see Eq. \ref{eq:n_vs_T}).}
    \label{fig:er_T}
\end{figure}

Furthermore, as shown in Figure \ref{fig:er_T}, the expected detection numbers increase toward the fainter end. This indicates that if the search can be extended to fainter stars, more FFP events could be detected. We find that a quadratic function of TESS mag $T$ can well fit the logarithm expected numbers,
\begin{equation}
    \frac{d N_{\rm FFP,S61}}{dT}\propto 10^{1.2065 T -0.0359T^2}.
    \label{eq:n_vs_T}
\end{equation}
Therefore, we extend the function to $T=16$ and integrate over the magnitude range to roughly estimate the yields for fainter stars. The expected yield is then increased by a factor of $N_{\rm FFP,S61}(T<16)/N_{\rm FFP,S61}(T<13.5)\approx4.2$ compared to the current bright star searches. 

Combining with the improvement from the new strategy, we extrapolate the yields to the entire first 7-yr TESS mission and find
\begin{equation}
    N_{\rm FFP,7yr}(T<16) \approx 1.5.
\end{equation}
However, this must be an upper limit because a search toward fainter stars must consider blending effects, which are not included in the above estimation.

\HL{In conclusion, we find that compare to the results presented in Section \ref{sec:res},}
there is still room for the optimization of the search strategy and the exploration of fainter targets. After applying these possible improvements, $\sim 1$ FFP detections are expected in the entire TESS database.
Therefore, TESS is still inefficient in finding microlensing FFPs. This highlights the necessity of future space-based FFP surveys, such as the planned Roman and Earth 2.0, which are expected to detect $\mathcal{O}(10^3)$ FFP events and significantly improve our understanding of FFPs.

\section*{}
All the {\it TESS} Input Catalog data used in this paper can be found in MAST \citep{TICdoi}. 

\acknowledgments
H.Y., W.Z., T.G., R.K., and S.M. acknowledge support by the National Natural Science Foundation of China (Grant No. 12133005). W.Z. acknowledges the support from the Harvard-Smithsonian Center for Astrophysics through the CfA Fellowship. The authors acknowledge the Tsinghua Astrophysics High-Performance Computing platform at Tsinghua University for providing computational and data storage resources that have contributed to the research results reported within this paper.

\bibliography{Zang.bib}

\end{CJK*}
\end{document}